\documentclass[useAMS,usenatbib]{mn2e}
\usepackage[]{graphicx}
\usepackage{psfig}
\usepackage{epsfig}
\usepackage{fixltx2e}

\title[Focal ratio degradation in lightly-fused hexabundles]{Focal ratio degradation in lightly-fused hexabundles}

\author[J. J. Bryant et al.]{J. J. Bryant$^{1,4}$\thanks{E-mail: jbryant@physics.usyd.edu.au (JJB)}, J. Bland-Hawthorn$^{1,2}$, L. M. R. Fogarty$^{1}$,  J. S. Lawrence$^{3}$ 
\newauthor 
 and S. M. Croom$^{1,4}$\\
$^{1}$ Sydney Institute for Astronomy (SIfA), School of Physics, The University of Sydney, NSW 2006, Australia \\
$^{2}$ Institute of Photonics \& Optical Science, The University of Sydney, NSW 2006, Australia \\
$^{3}$ Australian Astronomical Observatory, PO Box 915, North Ryde, NSW 1670, Australia\\
$^{4}$ ARC Centre of Excellence for All-sky Astrophysics (CAASTRO); \\
}

\begin{document}
\date{}
\pagerange{\pageref{firstpage}--\pageref{lastpage}} \pubyear{2011}
\maketitle

\label{firstpage}

\begin{abstract}

We are now moving into an era where multi-object wide-field surveys, which traditionally use single fibres to observe many targets simultaneously, can exploit compact integral field units in place of
single fibres.
Current multi-object integral field instruments such as SAMI \citep{Cro2012, JB2012a} have driven the development of 
new imaging fibre bundles (hexabundles)
for multi-object spectrographs. We have characterised the performance of hexabundles with different cladding thicknesses and compared them to that of the same type of bare fibre, across the range of fill-fractions and input f-ratios likely in an IFU instrument.
Hexabundles with 7-cores and 61-cores were tested for focal 
ratio degradation (FRD), throughput 
and cross-talk when fed with inputs from F/3.4 to $>$F/8.
The five 7-core bundles have
cladding thickness ranging from 1 to 8$\mu$m, and the 61-core bundles have
5$\mu$m cladding. As expected, the FRD 
improves as the input focal ratio decreases. We find that the FRD and throughput of the cores in 
the hexabundles match the performance of single fibres of the
same material at low input f-ratios. 
The performance results presented can be used to set a limit on
the f-ratio of a system based on the maximum loss allowable for a planned 
instrument.  Our results confirm that hexabundles are a successful alternative for fibre imaging devices for multi-object spectroscopy on wide-field telescopes and have prompted further development of  hexabundle designs with hexagonal packing and square cores.

\end{abstract}


\section{Introduction}
\label{intro}

Over the past two decades, single-fibre multi-object spectroscopic (MOS) surveys have amassed large galaxy
samples from which global properties and evolutionary trends have been deduced. However,
a fixed angular-sized aperture fibre can give misleading results when the 
same sized fibre is used to observe all galaxies irrespective of their size, distance 
or morphology (see for example, \citet{Ell2005}). Spatially-resolved spectroscopy is the way forward for future galaxy surveys and will lead to significant advances in our understanding of galaxies' morphologies and 
evolution. While integral field units (IFUs) have been very effective in 
studies of individual galaxies, up until recently, the number of objects that can be observed simultaneously is limited. Our motivation was to develop a technology that in the future can give spatially-resolved spectra of hundreds of galaxies across a field by replacing single fibres in multi-object robotic-positioners, with compact IFU devices.

To enable a large galaxy survey with resolved spectroscopy, we have 
developed imaging fibre bundles called {\it hexabundles} \citep{JBH2011}. 
Earlier hexabundle designs were constructed by strongly fusing the fibres. This
distorted the fibres, removing the interstitial holes, but at the same time
significantly worsened their optical performance \citep{JB2011}. These
bundles have been now superseded in preference for lightly-fused bundles 
in which the cores remain circular and have significantly better optical 
performance at the cost of a lower fill-fraction.

One of the key performance criteria for astronomy is to minimise focal
ratio degradation (FRD). FRD increases the output cone half-angle $\theta$ (where $NA=sin \theta$, and  f-ratio$\sim 1/(2NA)$) of light from
the optical fibre compared to the cone half-angle of light put in (also known as NA upconversion). The main causes of
FRD are due to light scattering 
in the fibre from irregularities and microbends, distortion of 
the fibre from stress, compression or tight bend radii and the quality of the fibre end finish \citep[for a detailed 
discussion of
the causes of FRD, see for example][]{Hay2011,Oli2005,Car1994}. 
The implication of FRD is that the spectrograph either 
needs to 
be physically bigger to collect all the light from the larger angles, or there will be
light lost from the system where the acceptance angle of the spectrograph is
exceeded. In order to put constraints on the applications of hexabundles, 
we have characterised the FRD of a number of
hexabundles in detail, using a range of input beam speeds used in astronomy.
The hexabundles have a range of cladding thicknesses, from 1--8$\mu$m and 
either 7 or 61 cores.
The hexabundle tests presented here were all using AFS105/125Y fibre. The 61-core hexabundles were then employed in the Sydney-AAO Multi-object Integral field spectrograph (SAMI) \citep{Cro2012, JB2012a} prototype version, before it was more recently upgraded to a new fibre type with new hexabundles.

In Section~\ref{setup} we outline the testing method and describe the specifications 
of the hexabundles being testing. The performance of the hexabundles in 
terms of FRD, throughput and cross-talk 
is discussed in Section~\ref{ResDis}. Section~\ref{AppFut} discusses future development and then the summary is given in Section~\ref{summary}.

\section{Description of the hexabundle devices and testing method}
\label{setup}

\subsection{Specifications of the hexabundles tested}

\begin{figure}
\centerline{\psfig{file=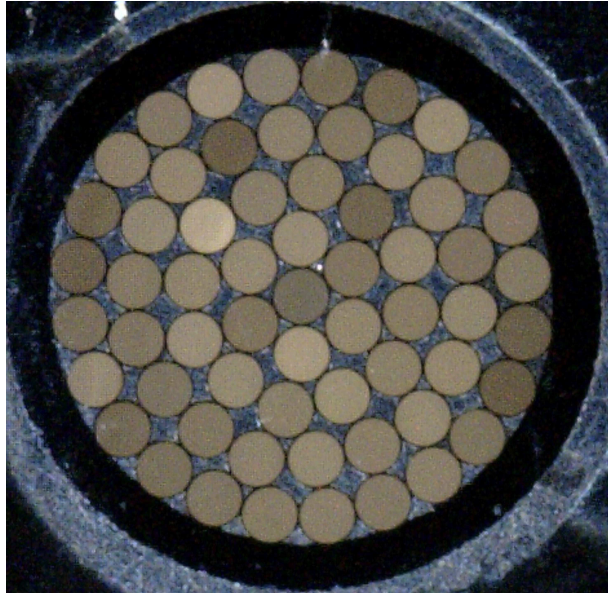, width=8.0cm}}
\vspace*{3mm}
\caption{
One of the 61-core lightly-fused hexabundles.
The interstitial holes are filled with
soft, low refractive index glue.
The cores are 105$\mu$m in diameter and 115$\mu$m with cladding. 
}
\label{bundle_pic}
\end{figure}

Five 7-core hexabundle devices were tested, each with a different cladding 
thickness, as well as two 61-core hexabundles (shown in Fig.~\ref{bundle_pic}).
The 7-core hexabundles have cladding thicknesses of 1, 2, 4, 6 and 8$\mu$m, 
while the 61-core hexabundles both had 5$\mu$m cladding thickness for all cores. 
The fibres used in each of the bundles was low-OH AFS105/125Y with a limiting numerical aperture (sine of the maximum acceptance cone angle) of 0.22 
and core and cladding diameter of 105$\mu$m and
125$\mu$m respectively, before the cladding was etched away over a $\sim2$\,cm length where the
fibres were then fused.
Each of the bundles has cores which are lightly fused so that the fibres
are mostly circular but interstitial
holes remain, giving fill-fractions (area of cores to that of the bundle) of 
between 0.87 and 0.67 (for 1 to 8$\mu$m cladding thickness respectively).
Beyond the fused region the individual fibres coming out of the hexabundle device have the full
10$\mu$m cladding thickness.

\subsection{Method for testing FRD, cross-talk and throughput}
\label{sec:method}

\begin{figure*}
\begin{minipage}[]{0.9\textwidth}
\centerline{\psfig{file=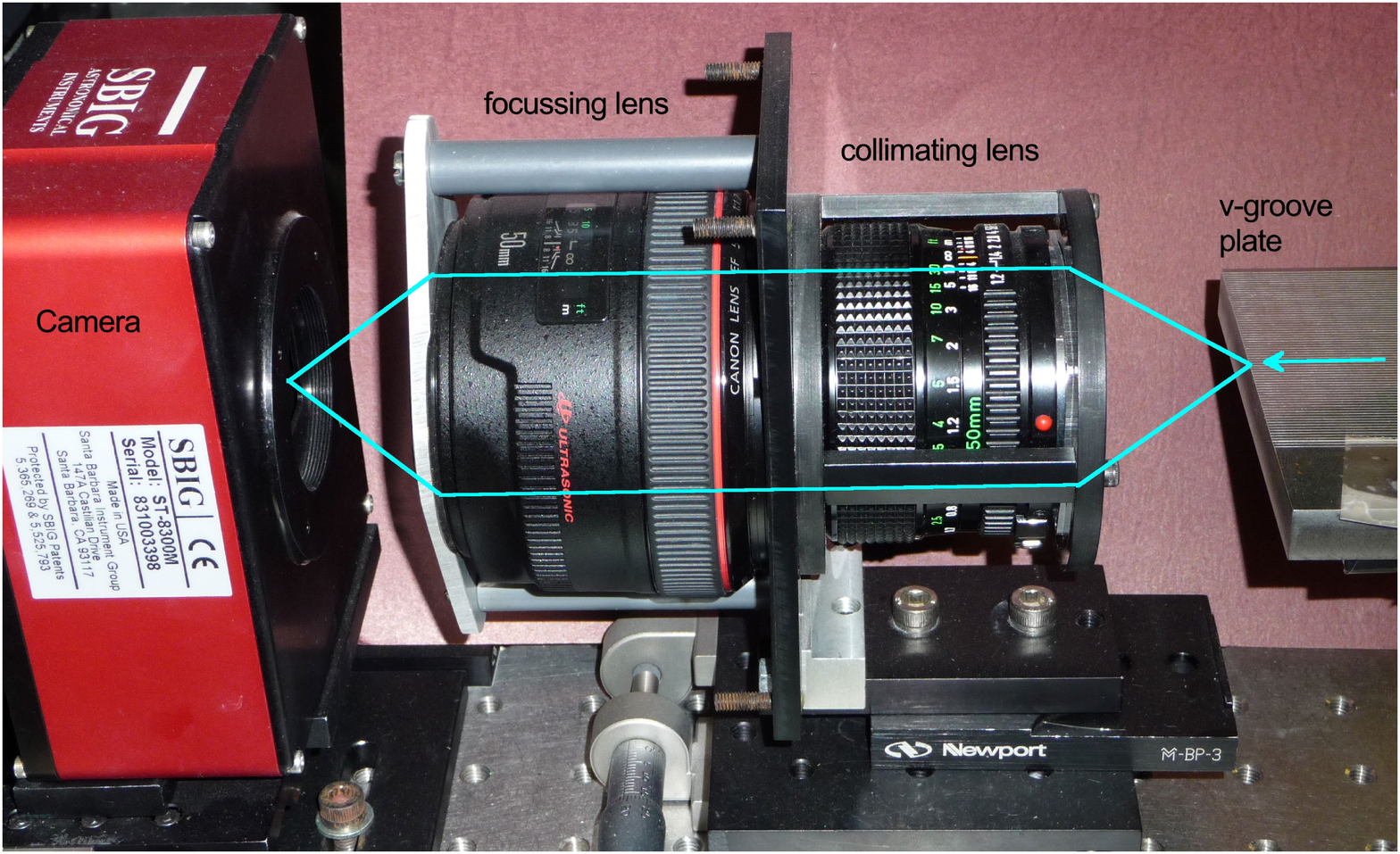, width=10.0cm}}
\end{minipage}%
\vspace*{0.1mm}
\begin{minipage}[]{0.9\textwidth}
\centerline{\psfig{file=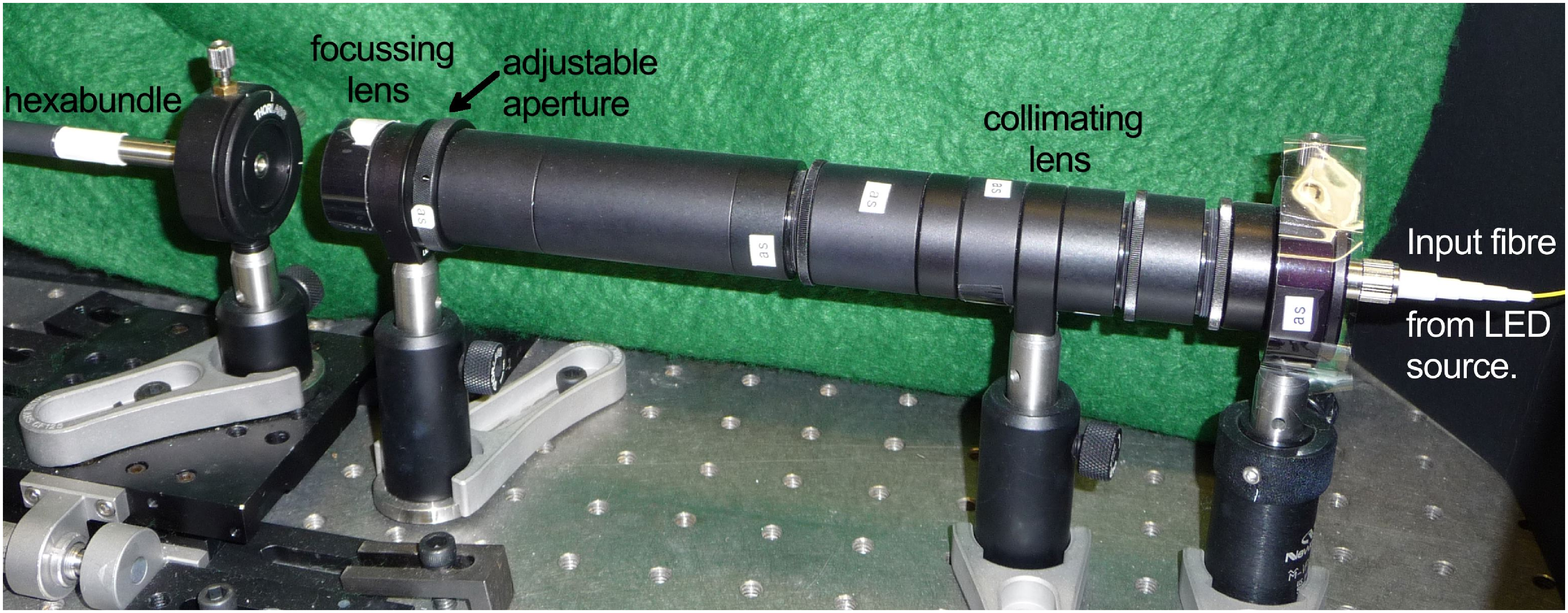, width=10.0cm}}
\end{minipage}%
\vspace*{3mm}
\caption{The testing set up for measuring FRD as described in section~\ref{sec:method}. The input system (lower) re-images an SMF-28 fibre to form an input light cone into the hexabundle (on the far left of the lower image) with an input NA defined by the adjustable aperture. The output system (top), re-images the fibre onto the camera.
}
\label{testbench}
\end{figure*}

Fig.~\ref{testbench} shows the setup used to assess FRD for all of the 
hexabundles tested. 
An Oriel LED source was focussed through a filter into an SMF-28 fibre which is 
multimoded in the wavelength range being tested, and has an apparent core size of $\sim10\mu$m. Two different filters were used in turn, for each 
measurement, and they were Bessell $B$ and $R$-band filters; the $B$-band filter centred on 457\,nm with a FWHM of 27\,nm and the $R$-band filter centred on $\sim 596$nm with an asymmetric profile of 60\,nm width.  The output of this fibre was aligned with the focus of a lens 
with a 75\,mm 
focal length. 
An adjustable iris in the collimated beam was used to set the beam speed
into the hexabundle within a range of F/3.4 to F/20. The profile
of the beam going into the iris was designed to significantly overfill the
iris so that the collimated beam fed through the iris to the focussing
lens was just the central part of the gaussian beam from the input
fibre. The result was close to a flat-topped square input function into
the hexabundles (see the input curves in 
Fig.~\ref{NAvsEE1} later).  
The collimated beam was then refocussed by 
a 35\,mm focal 
length lens to form the input beam which was fed into the hexabundles. The focussed F/3.4 input spot diameter was $50\mu$m (at $0.5$\% of the peak profile intensity) and therefore smaller than the core size 
to ensure no light was input into adjacent cores to contaminate cross-talk measurements.
Initially an SBIG camera was used to image the input beam in the far-field. The
camera was then replaced by a mount for the fused input end of the hexabundles, which was carefully aligned (see section~\ref{sec:errors} below)
on axis so each hexabundle could be mounted and measured in turn. 

Single fibres coming out of the hexabundles were cleaved (details of the cleave quality are discussed in section~\ref{frd} below) and mounted on a 
plate that has parallel v-grooves. Each fibre was gently secured to the plate
with tape, using the minimum pressure required to stop them from slipping, in 
order to reduce any stress on the fibre. The contribution of this method to the FRD is assessed in  section~\ref{sec:errors}.  The v-grooves were aligned with a 
pair of camera lenses, which  
collimated and then refocussed the output light.  An SBIG camera imaged the
output at a precise back focus position. 

The images were dark subtracted using separate dark images, then the barycentre of the imaged spot
was fitted. Encircled energy was calculated in concentric rings about the
centre position using aperture photometry packages within {\sc iraf} \citep{Tod1986}. Within the range of f-ratios being tested, both the input and output cone angles were very much less than the limiting NA of the fibre. We therefore use `NA' to describe $sin \theta$ of the cone half-angle $\theta$. Encircled energy is measured in increasing radii on the image. The radii, when combined with the back focal distance and pixel size then gives the cone half-angle and hence output NA versus encircled
energy was calculated. 
 Total integrated counts were then compared between the
core with the input light, and the surrounding adjacent cores, to assess cross-talk.

\subsubsection{Errors in FRD due to alignment of the optics and positioning of the fibres}
\label{sec:errors}

Any FRD test setup will have alignment uncertainties that lead to geometrical FRD which adds to the measured FRD of the fibre. Therefore accurate characterisation of the system errors is essential. The accuracy of the NA measurements from which FRD is assessed is significantly dependent on the
alignment of the testing apparatus, and the cleave on the ends
of the fibres. 

There are three main contributions to the measurement uncertainties in the {\it input}
NA or f-ratio feeding the hexabundle, and they are due to alignment. 
Firstly, the NA input into the hexabundle was measured using a camera in place of the
hexabundle holder. The camera was positioned on-axis on the stage used to focus the hexabundles. Any angle between the camera and the input beam will 
record an incorrect input NA and is apparent from fitting the spot image over the travel of the stage. The uncertainty in the input NA measurements due to the
alignment of the camera to the input beam is $\pm0.0005$.
Secondly, positioning the fibre holder relative to the input beam introduces an uncertainty
in NA into the fibre of $<\pm0.001$. Thirdly, the accuracy with which the adjustable aperture can be set, results in an NA error of $<\pm0.0005$.

The alignment uncertainties of the {\it output} optics further contribute to uncertainties in the NA measurements in the following way.
The V-groove plate was aligned with the optical axis, however, there are
slight angle differences depending on each fibre. This is due to {\it both} 
how the 
fibres sit in the v-grooves and potentially any bending of the output light
by sub-1-degree variations between cleave angles from fibre-to-fibre. 
These errors were 
measured by shifting the V-groove plate back and forward and imaging the shift
in each core. The resulting measured NA uncertainty was up to $\pm0.004$. To test the impact  of the tape holding the fibres on the plate, we applied tape firmly with repeated applications and found a maximum variation
in NA of less than $\pm0.002$. When lightly secured, the measurement uncertainty will therefore be $<<\pm0.002$. 
Alignment of the camera was found to give an NA uncertainty $<\pm0.001$. 
The focus on a $105\mu$m-core multimode fibre is less precise than for single-mode fibre as different modes focus at slightly different points. Coupled with any focussing errors from the optics, this results in 
an uncertainty in the NA of $\pm0.0067\times$NA, or $\pm0.0007$ to $\pm0.0013$ for an output
NA of 0.1 to 0.2 respectively.

Table~\ref{SummarytableFRD} summarises the errors. All uncertainties were combined in quadrature to give the measurement error, and
then the profile fitting errors were included to give the total uncertainties
listed in each figure (used in the calculation of output NA, NA upconversion and f-ratio errors)

Time variation in the input light intensity plus variations in the
SBIG camera response, were quantified with repeated images through one
core of the bundle in time periods ranging from second to hours. The 
maximum variation in resulting integrated counts was $<1.0$\% with typical
values of $\sim0.2$\%. 

It is important to note that these alignment errors in any FRD test setup can introduce geometrical FRD, which can worsen the apparent measured FRD. In that sense, the measured FRD from different setups will vary, and will always be a worst case. FRD results are therefore most meaningful when compared to a bare fibre measured with the same apparatus. 

\begin{table}
\caption{Summary of the contributions to uncertainties in FRD from measured NA errors along with other percentage errors affecting throughput. All values are the maximum (not typical) uncertainties.}
\label{SummarytableFRD}
\begin{center}       
\begin{tabular}{lc}
\hline
Alignment component & $\pm$ max. error in NA  \\
\hline
\hline
Input into hexabundle:  & \\
\hline
 Input beam camera alignment & 0.0005 \\
 Fibre holder positioning & 0.001 \\
 Input aperture repositioning & 0.0005 \\
\hline
Output reimaging optics:  & \\
\hline
 V-groove plate alignment & 0.004 \\
 Securing fibre to the plate & 0.002\\
 CCD camera alignment & 0.001\\
 End face focus & 0.0067$\times$ NA \\
 Chromatic aberrations in lenses & 0.0009 \\
\hline
\hline
Other uncertainties: & (\%)\\
\hline
Light source variability & $<1$ \\
Throughput from coupling position & $<<5$ \\
\hline
\end{tabular}
\end{center}
\end{table}

\subsubsection {Coupling position}

If the focussed input spot is centred on a fibre, more light will couple into the central modes. However, if the input spot is significantly offset from the centre, the amount of light that gets into the fibre will be different depending on the fraction of light coupled into higher-order modes.
The sensitivity of the results to the coupling position of the light into each core of the hexabundles was tested by aligning the input spot with
the centre of the fibre (using a microscope) and then offsetting it from the centre
to a number of positions, but still keeping the spot size entirely
within the core (the spot size was 
$<50\mu$m at 0.5\% of the peak 
intensity).
The resulting variation in throughput was measured as the integrated counts from aperture
photometry 
on the output ccd image, when the same light level was input. The input LED light source 
intensity varies by 1\%, while the throughput varied by up to 5\%. This
agrees with the results of \citet{Hor2007} who similarly
found a variation of only 5\% in throughput between different coupling
positions within the core of multi-mode fibres.  The same test on a hexabundle core
found the same 5\% variation in throughput due to coupling. 
The precision with which we could align to the centre of the fibres using the microscope, allowed a smaller centring uncertainty than the range of coupling positions tested, and we therefore expect the throughput error from coupling to be $<<5$\%.

\section {Results and discussion}
\label{ResDis}

\subsection{Focal Ratio Degradation (FRD)}
\label{frd}

\begin{figure*}
\hspace*{4mm}
\begin{minipage}[]{0.55\textwidth}
\centerline{\psfig{file=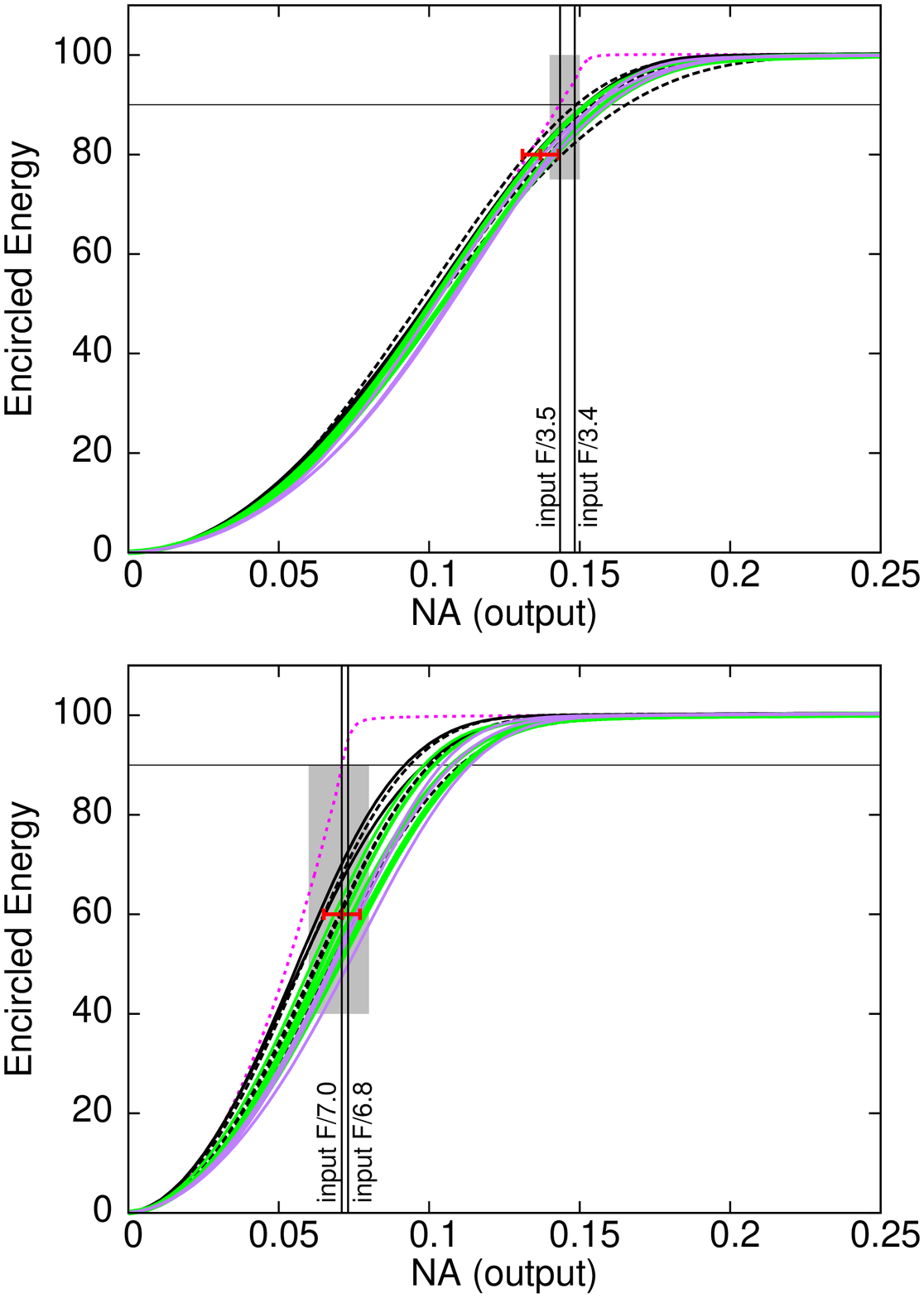, width=11.0cm}}
\end{minipage}%
\hspace*{0.1mm}
\begin{minipage}[]{0.40\textwidth}
\centerline{\psfig{file=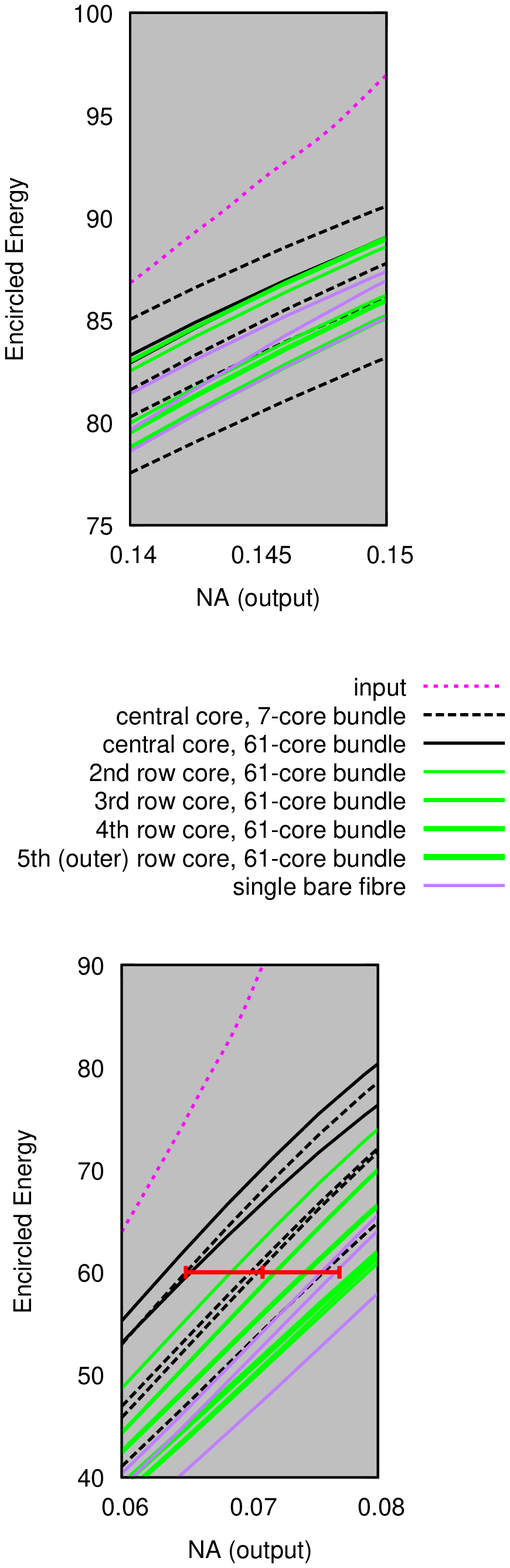, width=10.0cm}}
\end{minipage}%
\caption{Encircled Energy versus Numerical Aperture (NA$=sin\theta$; for output cone half-angle $\theta$) 
profiles through the Bessel blue ($\sim$0.45 $\mu$m)
filter for
the central cores (black) of the 7-core (dashed lines) and 61-core (solid 
lines) hexabundles as well as a number of the non-central cores from a 
61-core hexabundle (green lines).  The thickness of the green lines is proportional to how far the core is from the centre of the bundle. Grey shaded box regions are shown magnified on the right.
The dotted magenta line
is the input beam profile. The NA uncertainty due to alignment, focus and measurement
errors is $\pm 0.006$ as illustrated by the representative red error bar.
For clarity, plots are only shown for two of the input f-ratios tested -  when 95-90\% of the encircled energy of the input light cone was within F/3.4-3.5 (top row), and F/6.8-7.0 (bottom row), and the equivalent NA positions for these f-ratios are marked by the vertical black lines.
Purple lines are the curves for three single bare fibres that have the 
same AFS105/125Y fibre as the bundles.  At low input F-ratio, the error bar is larger than the scatter between curves.
} 
\label{NAvsEE1}
\end{figure*}

The impact of FRD on the encircled energy curves is shown in Fig.~\ref{NAvsEE1}. Worse FRD shifts the curve
to the right because the output cone angle $\theta$ is larger, giving a larger NA for a given  percentage of the encircled energy. The horizontal 
gap between the input and output profiles at 90 or 95\% encircled energy is the NA upconversion. Worse FRD corresponds to a larger NA upconversion. For clarity, we have shown a representative range of curves, highlighting the difference between the central cores and outer cores in several typical hexabundles.  
As with all fibres, the FRD of the AFS105/125Y fibre is worse when the input f-ratio is higher (Fig.~\ref{NAvsEE1}, lower row). At low f-ratios (F/$\sim$3.4, top row), all the fibres tested 
demonstrated the same FRD within errors.

Repeatability of these profiles
was checked by reimaging the 61-core bundle after re-centring the input
spot, refocussing, changing the input light
levels, and refitting the profiles. Over 16 images taken of one core with these changes, the output NA versus encircled
energy profiles were coincident within an NA scatter of
only $\pm0.003$. 

\begin{figure*}
\begin{minipage}[]{0.3\textwidth}
\centerline{\psfig{file=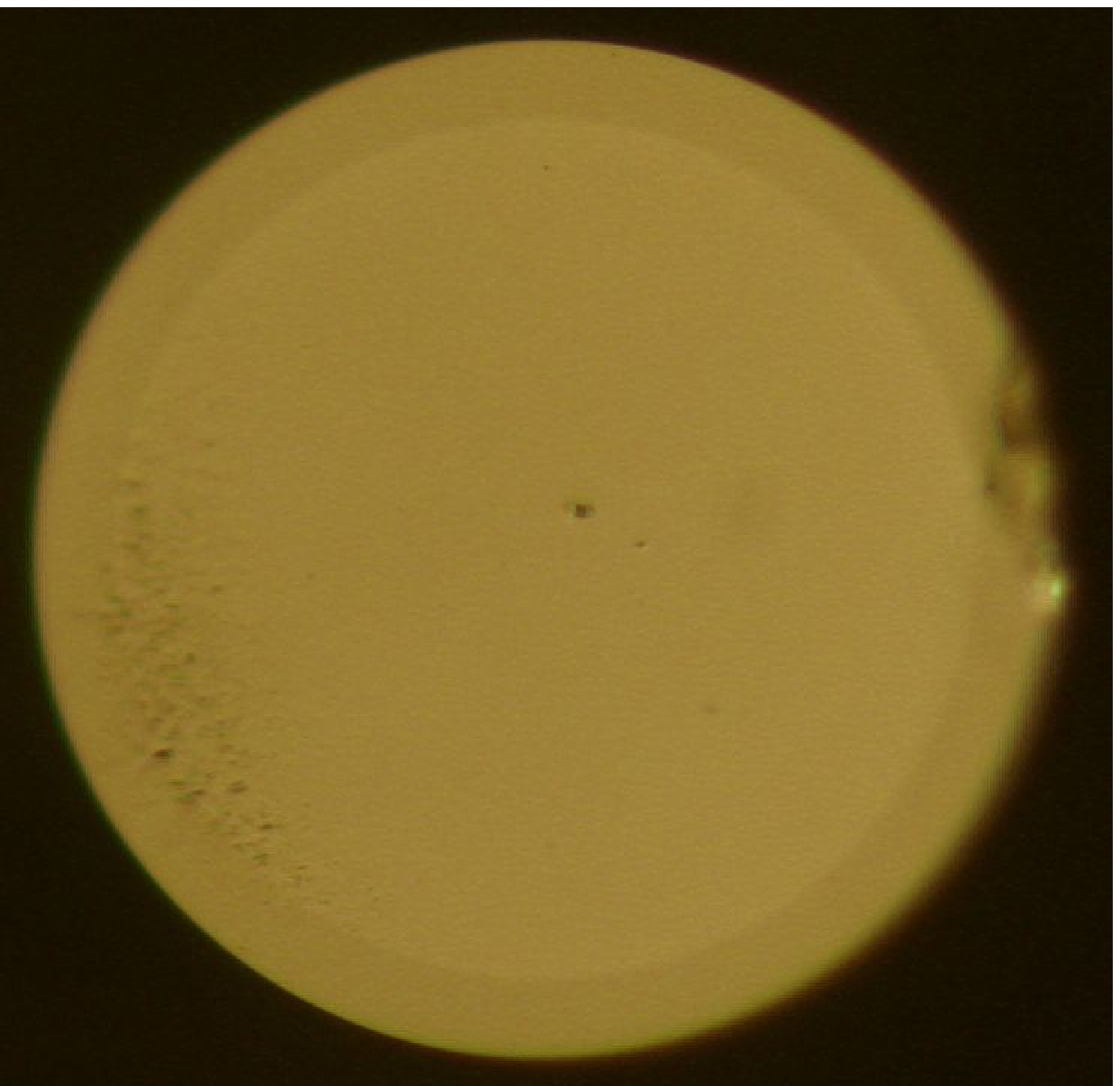, width=5.0cm}}
\end{minipage}%
\hspace*{0.1mm}
\begin{minipage}[]{0.3\textwidth}
\centerline{\psfig{file=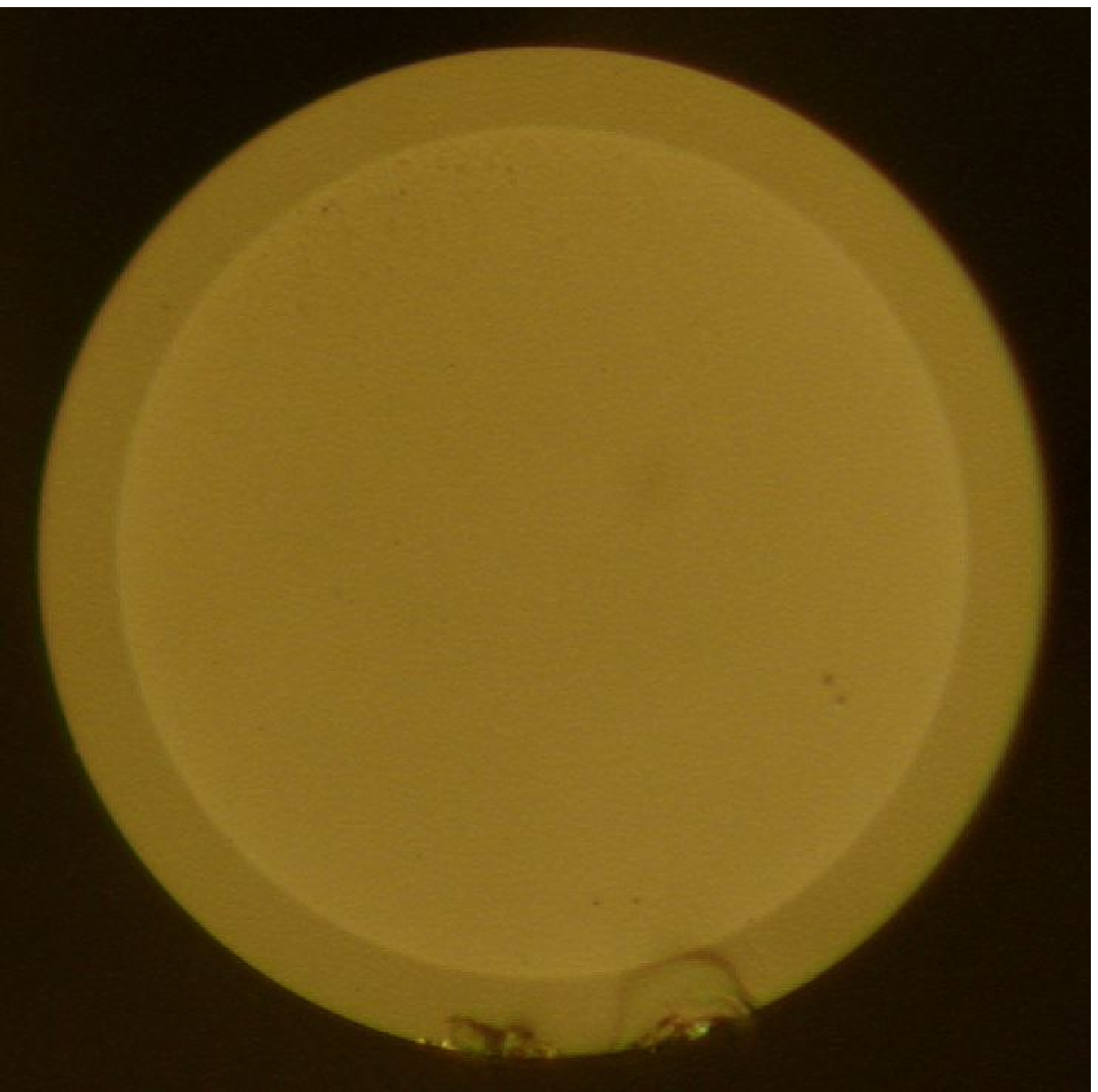, width=5.0cm}}
\end{minipage}%
\hspace*{0.1mm}
\begin{minipage}[]{0.3\textwidth}
\centerline{\psfig{file=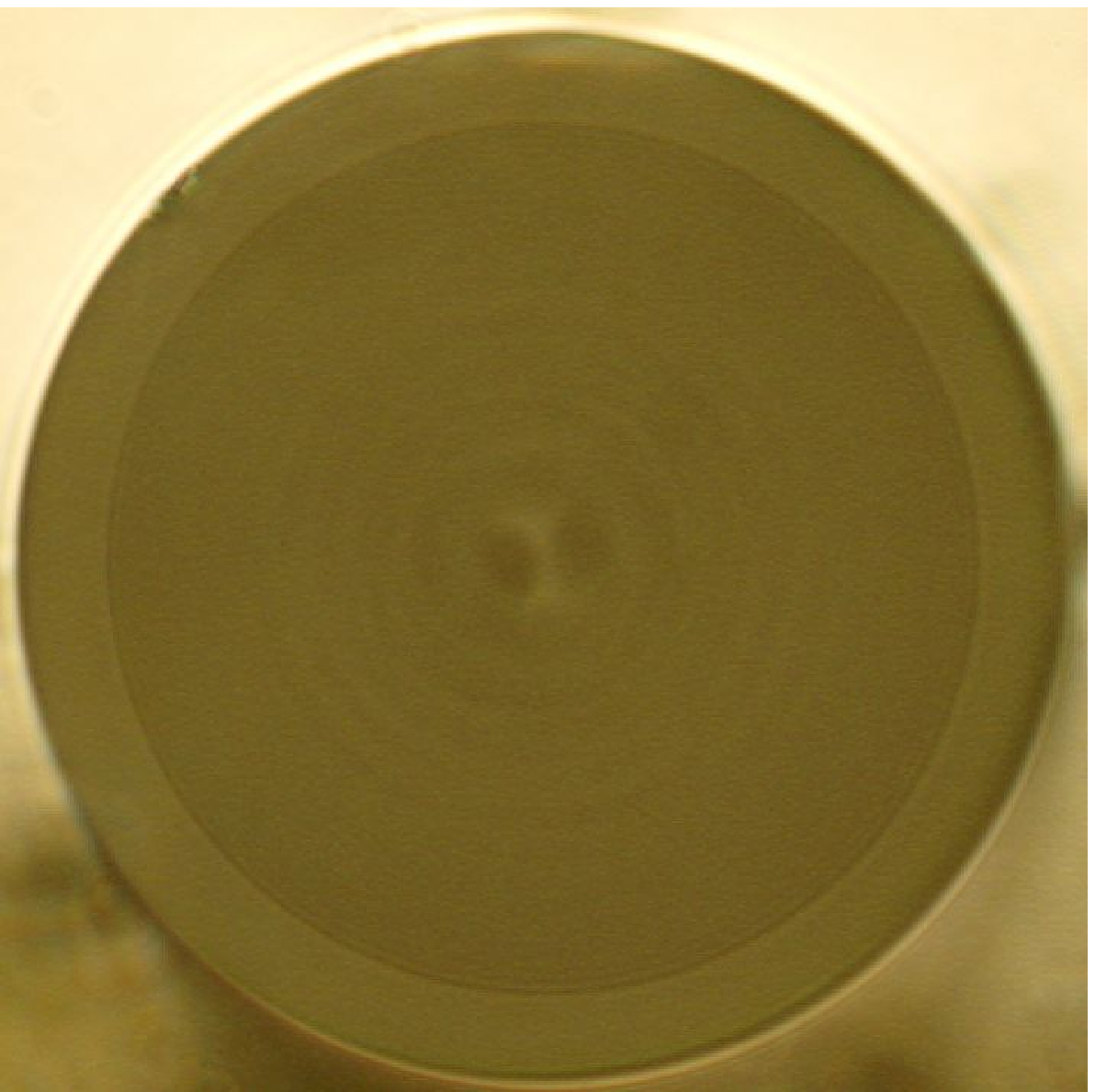, width=5.0cm}}
\end{minipage}%
\vspace*{3mm}
\caption{Microscope pictures of individual cleaved fibres at the output of the hexabundles. Typical unacceptable surface 
structure seen on cleaved fibre cores include a rougher surface across
part of the core opposite the cleave impact point, as well as individual 
pits in the surface (both shown in left image). Fibres were re-cleaved to
minimise these defects, to give typical core-finish shown in the centre and
right images. Fibres were also re-cleaved if the cleave impact 
reach the core as in the central image. The right image illustrates the finish we considered acceptable.} 
\label{microscope}
\end{figure*}

Variability in the FRD curves can potentially come from the
end cleave. 
The cleaves on the output of individual fibres were very carefully checked under a 
microscope to ensure that the cleave angle was $<1$ degree and that there
were no cracks or damage to the core. We ensured that the score mark  
from the cleave was only in the cladding and not near the core and there were
no scratches or residual damage from the cleave on the core. In each case
there was some surface roughness on the core which we attempted to minimise.
The surface roughness was gauged from the microscope images, some examples
of which are shown in Fig.~\ref{microscope}. We tested FRD of cores with
a range of apparent surface features after many re-cleaves of the same cores.
We found a point at which the remaining surface features did not change the output encircled energy profiles within the errors. We therefore believe that the difference in encircled
energy measurements between cores is not dominated by variations in the cleaves
for the 61-core bundles. 
This is further confirmed by the reduced scatter at lower input f-ratio. If the
cleaves strongly affected the profiles then we would see the same scatter 
between curves as we do at higher input f-ratios.

\begin{figure*}
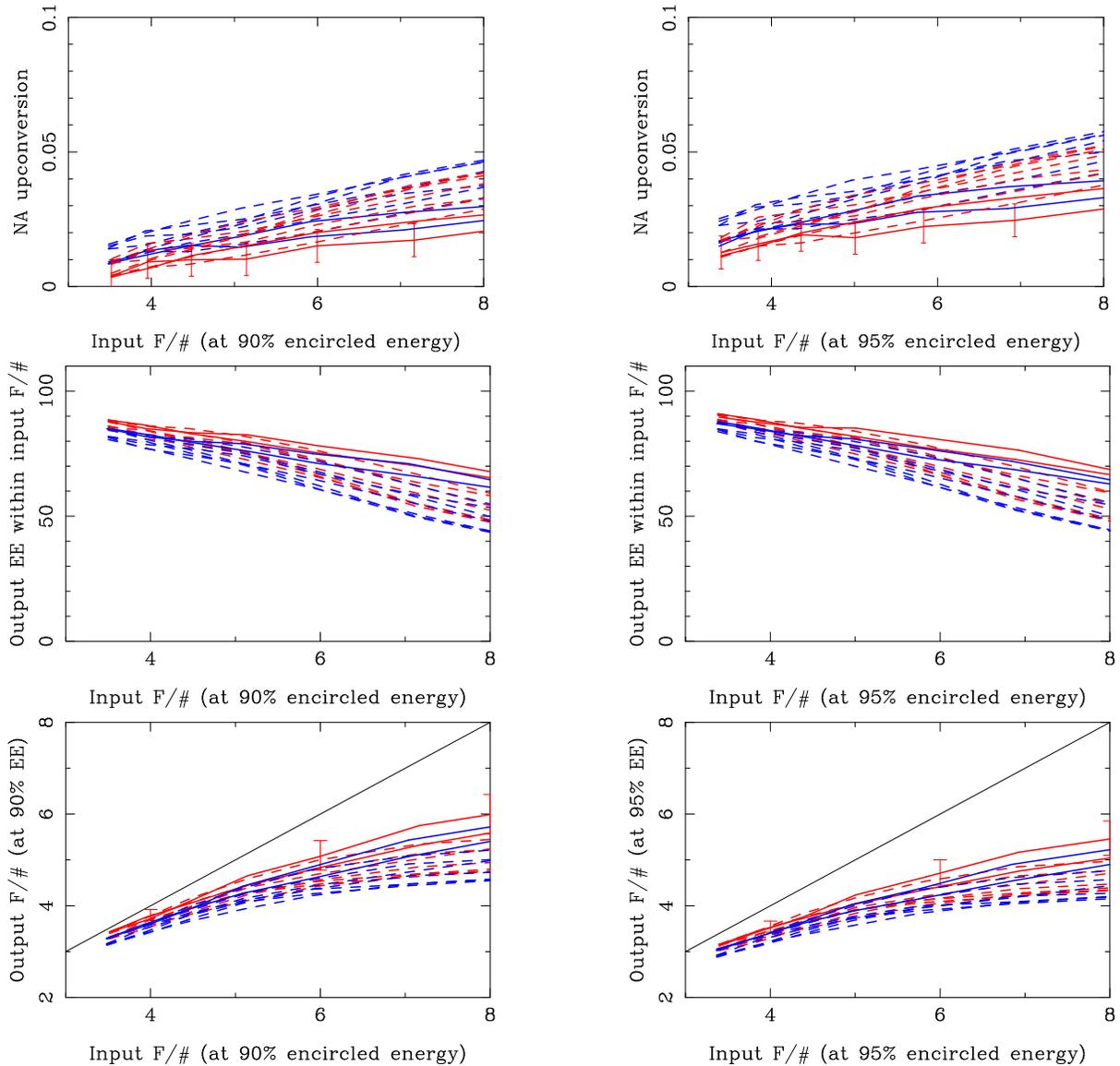

\begin{minipage}[]{0.47\textwidth}
\centerline{\psfig{file=FinvsNAupconv_90.ps, width=5.0cm, angle=-90}}
\end{minipage}%
\hspace*{5mm}
\begin{minipage}[]{0.47\textwidth}
\centerline{\psfig{file=FinvsNAupconv_95.ps, width=5.0cm, angle=-90}}
\end{minipage}%
\vspace*{1mm} 
\begin{minipage}[]{0.47\textwidth}
\centerline{\psfig{file=FinvsEEwithinFin_paper90.ps, width=5.0cm, angle=-90}}
\end{minipage}%
\hspace*{5mm}
\begin{minipage}[]{0.47\textwidth}
\centerline{\psfig{file=FinvsEEwithinFin_paper95.ps, width=5.0cm, angle=-90}}
\end{minipage}%
\vspace*{1mm} 
\begin{minipage}[]{0.47\textwidth}
\centerline{\psfig{file=Fin_Fout_paper_90.ps, width=5.0cm, angle=-90}}
\end{minipage}%
\hspace*{5mm}
\begin{minipage}[]{0.47\textwidth}
\centerline{\psfig{file=Fin_Fout_paper_95.ps, width=5.0cm, angle=-90}}
\end{minipage}%
\vspace*{1mm}
\caption {\small Focal ratio degradation results for the central core (solid lines) and outer cores (dashed lines) of two 61-core hexabundles. Each plot is shown for the case where the f-ratio is measured at 
90\% (left) and 95\% of the total encircled energy (right). Red and blue lines
are results through the Bessel $R$ and $B$ ($\sim$0.65 and 0.45$\mu$m) filters respectively.
{\it Top:} NA upconversion (difference between output and input NA or $sin\theta$ for half-cone 
angles $\theta$ at 90\% and 95\% encircled energy) versus the f-ratio of the input beam. For clarity, the error bars are shown
for one curve only, but they are representative of the uncertainties for each curve.
These
are the errors in the data for this core in this bundle, not the errors
due to the variance between cores and bundles.
{\it Centre:} Input f-ratio versus the percentage of encircled energy within the
same f-ratio at output.
{\it Lower:} Input f-ratio versus output f-ratio at 90/95\% encircled energy. Within errors, there is no FRD at $\sim$F/3 (90\% encircled energy), but FRD worsens with higher input f-ratios. 
} 
\normalsize
\label{NAup_cross}
\end{figure*}

It is notable from Figs.~\ref{NAvsEE1} (e.g. black solid lines) and~\ref{NAup_cross} (solid lines) that the central core of the bundles has better FRD than the
outer cores. When the input f-ratio is a low value, the difference is less than the 
errors. However, for higher input f-ratios where the difference is more apparent, we attribute the worse FRD of the outer cores to more distortion than the central
core due to the symmetry of the bundle. The outer cores are essentially circular,
but even the light fusing has a small effect which can worsen the FRD and is noticeable when the input f-ratio is high.

All FRD measurements were taken with cleaved cores. The hexabundles
may well perform better with polished ends or index matching gel coupling to a glass slide.
In current uses of these hexabundles (i.e. in the original SAMI instrument before upgrade) the loose 
hexabundle fibres were spliced to much longer fibre runs that terminate
in polished slit blocks. The aim was to retest the hexabundles through these
polished slit blocks, however the FRD introduced through this particular 
fibre run significantly degraded the FRD to make it an invalid
test of the hexabundle performance. We have therefore not got a direct
test of these hexabundles with polished fibres or with index matching gel and
glass slides. Comparisons of these end finishing techniques have been done
before by others and the results can be used to predict how much better
the hexabundles may perform with different end finishing. 
For example, \citet{Hay2011} (their table 4) found a 5\% increase in encircled energy
at F/3.6 (at 532nm) for index-matching gel plus glass slide compared to a good cleave.
This level of improvement in Fig.~\ref{NAvsEE1} (top) would 
shift the outer encircled energy profile for the worst FRD cores to agree with 
the best central cores, and the best cores would be better still.
\citet[][their fig. 5]{Pop2010} compared a polished end finish with the 
combination of a cleaved fibre plus index-matching gel. Across the range of 
input f-ratios we have tested, our central fibre results are similar to the 
polished fibre results in their fig.5 (both using 95\% encircled energies).
They show that cleaving and using index-matching gel improves the FRD somewhat.
Therefore we assume that the performance of the hexabundle fibres when used in
an instrument with index-matching gel will be at least as good as our 
central core results, but likely better, and the plots in 
Fig.~\ref{NAup_cross} should be interpreted with this in 
mind.

The FRD was first quantified by the difference in NA between the
output and input profiles at an encircled energy of 90 and 95\% (NA upconversion) and was 
measured for input beams speeds from F/3.4 to $>$F/20. The result  
is shown in Fig.~\ref{NAup_cross} (top) for F/3.4 to F/8. 
A second measure of FRD is how much light is lost from within the input cone angle by output. This is plotted as the percentage of the total encircled energy out of a fibre that is within the same light cone angle as the input (defined by the input f-ratio) as in Fig.~\ref{NAup_cross} (middle). The third numerical quantification of FRD comes from a measure of the output angle (or f-ratio) that contains 90 or 95\% of the encircled energy at each input f-ratio (Fig.~\ref{NAup_cross} lower). 
For an input beam at $\sim$F/3, the bundles have little or no 
FRD as the NA upconversion and output f-ratios are (within the error bars), in agreement with the 
input values. In fact the FRD shown for $\sim$F/3 includes the geometrical FRD in the test equipment within the errors as discussed in section~\ref{sec:errors}. Like all fibre systems, hexabundles lose less light from FRD when the input f-ratio is a low value (e.g. SAMI has F/3.4 input). However by
$\sim$F/8, the central fibres have $\sim64-69$\% of the light
from the fibre, coming out within F/8 (Fig.~\ref{NAup_cross}, middle).

In practice, fibres are used in astronomy with low input f-ratios because of the impact of FRD. It is less productive to use fibres beyond F/8, however, at any f-ratio, the light loss due to FRD can be minimised with a larger
acceptance cone into the spectrograph. For example, if F/6 is fed into
the hexabundle, then $>90$\% of the light will go into a spectrograph that
accepts an $\sim$F/5 beam for the best cores with good end finish (Fig.~\ref{NAup_cross}, lower left panel).

\vspace{-1mm}

While FRD has been shown to be dependent on fibre end finish, microbends,
and stresses on the fibre, the wavelength dependence of FRD is unresolved. 
Some studies have found no dependence \citep{Cra2008,Sch2003}, however \citet{Pop2007}
found worsening FRD with increasing wavelength, but not to the extent predicted by the theoretical models of \citep{Glo1972},
while \citet{Mur2008} found improving FRD with increasing wavelength. 
Our FRD measurements have an advantage that the two separate filters
were exchanged with no change to the fibre position, cleave, mount or 
input NA. Therefore, the effect due to wavelength is isolated from other
contributions to FRD. While the FRD at the two wavelengths overlaps within 
errors, the shorter wavelength is consistently worse 
for all fibres measured (see  
Fig~\ref{NAvsEE1} and ~\ref{NAup_cross}). Chromatic aberrations in the input lenses
have been ruled out as the cause because the effective input NA was measured
separately at each wavelength and the difference found to be $<0.0009$. 
However, the difference in input f-ratio for a given output f-ratio in Fig~\ref{NAup_cross} between colours is very much larger than this (e.g. for output 
of F/4.7 at 95\% encircled energy, the difference between the red and blue input f-ratio is equivalent to an NA or 0.005, 5 times that possible from 
chromatic aberrations). In order to confirm the extent of the 
wavelength dependence, a larger wavelength range would need to be tested.

\subsection{Comparison of hexabundle and single fibre results}

In order to assess how the hexabundles perform compared to a single fibre
of exactly the same type and length, bare AFS105/125Y fibres were mounted
into an SMA connector on one end and cleaved at both ends. They were then tested in
exactly the same way as the hexabundles. 
There were differences between batches of bare fibre, such that the bare fibre FRD curves spanned the same range of FRD as the curves for the central and outer hexabundle curves, but repeated samples from the same batch were always consistent within the errors. For clarity, only a sample of the bare fibre curves are shown in Fig.~\ref{NAvsEE1} (two from the same batch, and one from a different batch). While some of the bare fibres tested had similar FRD to the best central hexabundle cores, none had less FRD. Therefore the hexabundles perform as well as bare fibre.

\begin{figure}
\centerline{\psfig{file=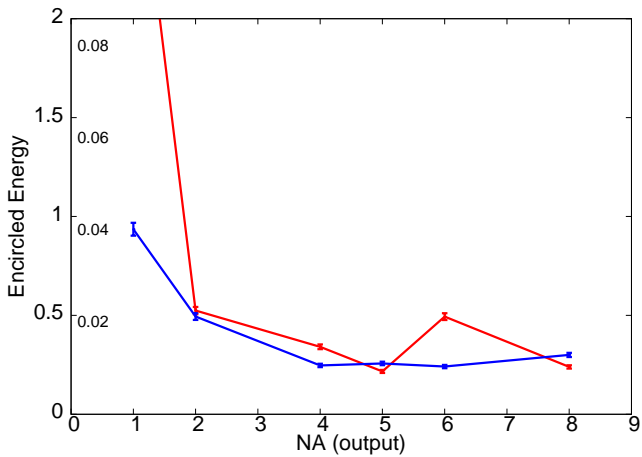, width=6.0cm, angle=-90}}
\vspace*{1mm}
\caption{
The \% cross-talk (also given in dB inside the box) from the
central fibre into all 6 surrounding fibres at F/3.4 input versus the cladding thickness of the fibres. The 7-core  hexabundles have 8, 6,
4, 2 and 1$\mu$m cladding, while a representative 61-core hexabundle has $\sim 5\mu$m cladding. The line colour represents the results through the $B$ and $R$-band filters. Notice that the cross talk is insignificant at below $\sim0.5$\% in all but the thinnest cladding. The error bars include the light source variability and throughput variations from input coupling position.
} 
\label{crossTfig}
\end{figure}

\subsection{Cross-talk}
\label{crossT}

In a hexabundle, scattering of light out of one core, may result in coupling into an adjacent core, which will then be seen as cross-talk. 
Thinner cladding allows more coupling of higher-order modes between cores. The measured
cross-talk is given in 
Fig.~\ref{crossTfig}.  The cross-talk is $<\sim0.5$\% for  2$\mu$m cladding thickness
and above. There is a marked increase in cross-talk for 1$\mu$m cladding, at which point the cross-talk outweighs the higher fill-fraction of 87\%. However, increasing the cladding to 
2$\mu$m, only decreases the fill-fraction by 3\%, while decreasing the cross-talk to $\sim$0.5\%.
For practical purposes the cross-talk measured here is very small compared to the effect of seeing. In expected uses of hexabundles, where each core is approximately the same angular size as the FWHM of the seeing disk 
(e.g. in SAMI), 45\% of the power from the seeing disk is in the adjacent cores 
(and 50\% in the central core).
In that case, 0.5\% cross-talk is negligible. The cores would need to be several seeing FWHM 
across for this level of cross-talk to be significant.

The cross-talk is slightly higher in the red filter compared to the blue. While the scattering model
of \citet{Glo1972} predicts that scattering should be worse at longer wavelengths we see a much
smaller effect than predicted by that model, and in fact have found the FRD to be worse in the blue. However, 
in the case of cross-talk, the much poorer blue throughput of the fibres most-likely has a larger
impact than scattering. This is because only light travelling at a large angle to the optical axis of the fibre will be lost from one fibre into
the next, and therefore the path length of that light will be larger by the time the light exits the adjacent
fibre. A larger path length will suffer more absorption in the blue leading to a smaller
measured output from the adjacent fibres and hence a smaller cross-talk in the blue.

\subsection{Throughput}

Throughput losses in individual fibres can be due to absorption in the core material or scattering of higher-order modes due to imperfections. 
Where that scattering leads to cross-talk, the cross-talk between fibres serves to give a loss from an individual 
fibre but not a systemic loss from the resultant image. Therefore, 
in the following throughput tests, any cross-talk (see Fig.~\ref{crossTfig}) will
have reduced the measured throughput by up to 1\%, however this loss would be
recovered in the image from a telescope.

\begin{figure*}
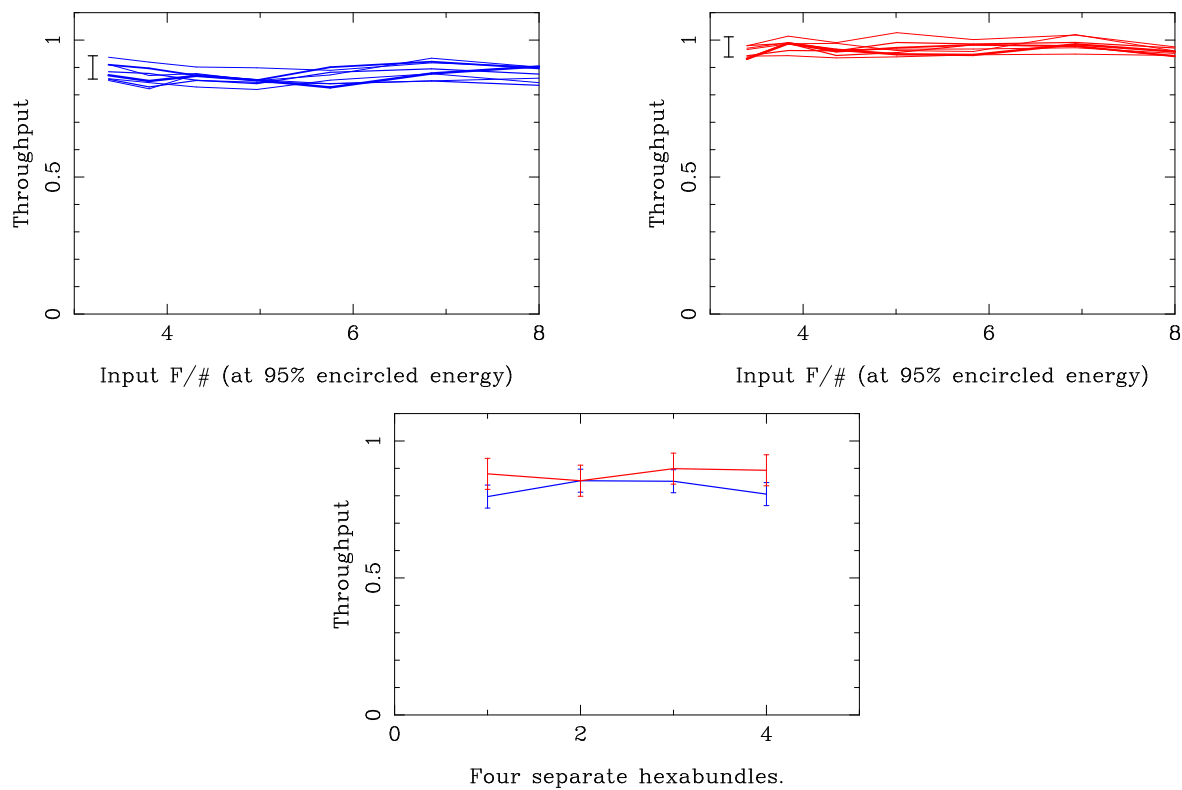

\begin{minipage}[]{0.47\textwidth}
\centerline{\psfig{file=thru_core2core_b2.61_b.ps, width=5.0cm, angle=-90}}
\end{minipage}%
\hspace*{0.3mm}
\begin{minipage}[]{0.47\textwidth}
\centerline{\psfig{file=thru_core2core_b2.61_r.ps, width=5.0cm, angle=-90}}
\end{minipage}%
\vspace*{3mm}
\centerline{\psfig{file=thru_bundle2bundle.ps, width=5.0cm, angle=-90}}
\vspace*{3mm}
\caption{Input f-ratio (at 95\% encircled energy) vs throughput of individual
fibres. The top two plots show multiple fibres within one hexabundle, through
the blue (left) and red (right) filters. Typical errors bars
for each point are shown on the left side of each plot. The lower plot shows the throughput
for a central fibre in four different hexabundles when the input was F/3.4.
The red and blue lines are through those respective filters.  
}
\label{thru}
\end{figure*}

Fig.~\ref{thru} compares the throughputs of several fibres within one bundle, 
as well as the central fibre in 4 different bundles. 
Data reduction techniques applied in aperture photometry were used to measure the total 
output counts (which is equivalent to the 100\% level in the encircled energy
plots in Fig~\ref{NAvsEE1}) in order to decouple the throughput losses from FRD losses. The ratio of the output to the input counts
was then corrected for the 3.3\% reflection at the air/glass interface at
either end of the bundle. This correction was necessary because in any 
astronomical instrument, the hexabundle face should be anti-reflection coated 
(and the test bundles were not), and the loose fibre ends will be  
spliced to fibres that feed directly into a spectrograph, typically coupled
in with index-matching gel. 
Small fluctuations in the LED source, input light photometry,  
photometric fitting errors, errors in the assumed value for the air/glass interface reflection, and repeatability of the aperture size setting, all contributed to the total error in this 
throughput test.

The throughput of the fibres relative to the input light is consistent
within errors, across all the input f-ratios tested. This is because at all the f-ratios shown, the effect of FRD is not substantial enough to fill the
maximum NA of the fibres (see Fig.~\ref{NAvsEE1}). Absorption is the dominant 
loss, and is responsible for the reduced throughput at shorter wavelengths in this AFS105/125Y fibre. It is notable that this loss was a driver behind upgrading the hexabundles and fibres used in the SAMI instrument, to a fibre type with higher throughput. 

Fibre-to-fibre throughput variations within a bundle 
are less than the errors giving a consistent imaging bundle.
Between 4 different bundles tested, the central core also showed throughput
variations consistent with the fibre-to-fibre variations, indicating that 
there is consistency in the batches of hexabundles.

\section{Applications and future development}
\label{AppFut}

The first on-sky demonstration of hexabundles was in the SAMI
instrument on the Anglo-Australian Telescope (AAT), which has now seen two generations of hexabundles, starting with the 61-core models from this paper. While a number of different circular-packed hexabundle types have been made, new developments of hexabundles
are focussed on increasing the fill fraction by using different packing geometries 
including a regular hexagonal packing of circular cores, or alternatively, using square core fibres. A
discussion of the trade-offs with these geometries can be found in \citet{JB2012b}.

The success of hexabundles has led to plans for a much larger robotically-positioned IFU instrument
called HECTOR \citep{Law2012}. In preparation for HECTOR, hexabundles are being tested in 
autonomous position robots called starbugs \citep{Gil2012}.

\section{Summary}
\label{summary}

Hexabundles have been developed with FRD and throughput performance at low input f-ratios, equivalent to that of the  bare fibre they are made from. Therefore, they can replace single fibres in multi-object spectroscopy, with the advantage of spatially-resolved spectroscopy on many objects simultaneously. 

In order to test the trade-offs between fill-fraction and cross-talk, hexabundles have been made with a range of cladding thicknesses. For cladding thicknesses of 2$\mu$m or more, the cross-talk has been shown to be negligible compared to the effects of seeing, while delivering fill-fractions as high as 84\%.
FRD performance improves with decreasing f-ratio, and using the plots shown, the viability of any fibre system with a larger input f-ratio can be assessed from the FRD losses. 

The early success of the hexabundle technology \citep{Cro2012,Fog2012} has convinced us that this approach will come to dominate future large galaxy surveys.

\subsection*{Acknowledgements}

We would like to thank Sergio Leon-Saval for useful discussions on topics surrounding this work. 

This research was conducted by the Australian Research Council Centre of Excellence for All-sky Astrophysics (CAASTRO), through project number CE110001020.


\begin{thebibliography}{}
\bibitem[Bland-Hawthorn et al.(2011)]{JBH2011} Bland-Hawthorn J. et al., 2011, Optics Express, 19, 2649
\bibitem[Bryant et al.(2011)]{JB2011} Bryant J. J., O'Byrne J. W., Bland-Hawthorn J., Leon- Saval S. G., 2011, MNRAS, 797
\bibitem[Bryant et al.(2012a)]{JB2012a} Bryant J. J., et al., 2012a, SPIE 8446, 31
\bibitem[Bryant et al.(2012b)]{JB2012b} Bryant J. J., Bland-Hawthorn J., 2012b SPIE 8446, 250
\bibitem[Carrasco \& Parry(1994)]{Car1994} Carrasco E., Parry I.R., 1994, MNRAS, 271, 1
\bibitem[Crause, Bershady \& Buckley(2008)]{Cra2008} Crause L., Bershady M., Buckley D., 2008, SPIE, 7014, 210
\bibitem[Croom et al.(2012)]{Cro2012} Croom S., et al., MNRAS 421, 872 (2012)
\bibitem[Ellis et al.(2005)]{Ell2005} Ellis S. C., Driver S. P., Allen P. D., Liske J., Bland-Hawthorn J., De Propris R., 2005, MNRAS 363, 1257
\bibitem[Fogarty et al.(2012)]{Fog2012} Fogarty L. M. R. et al., 2012 ApJ 761, 169
\bibitem[Gilbert et al.(2012)]{Gil2012} Gilbert J., et al., 2012, SPIE 8450, 14 
\bibitem[Gloge(1972)]{Glo1972} Gloge D., Bell. Syst. Tech. J., 151, 1767 (1972)
\bibitem[Haynes et al.(2011)]{Hay2011} Haynes D. M., Withford M. J., Dawes J. M., Lawrence J. S., Haynes R., 2011, MNRAS, 414, 253
\bibitem[Horton \& Bland-Hawthorn(2007)]{Hor2007} Horton A. \& Bland-Hawthorn J., 2007, Op Ex, 15, 1443
\bibitem[Lawrence et al.(2012)]{Law2012} Lawrence J., et al., 2012, SPIE 8446, 195
\bibitem[Murphy et al.(2008)]{Mur2008} Murphy J. D., MacQueen P. J., Hill G. J.,Grupp F., Kelz A., Palunas P., Roth M., Fry A., 2008, SPIE, 7018, 92 
\bibitem[Oliveira, de Oliveira \& dos Santos(2005)]{Oli2005} Oliveira A.C., de Oliveira L.S., dos Santos J.B., 2005, MNRAS, 356, 1079
\bibitem[Poppett \& Allington-Smith(2010)]{Pop2010} Poppett C. L., Allington-Smith J. R., 2010, MNRAS, 404, 1349
\bibitem[Poppett \& Allington-Smith(2010)]{Pop2007} Poppett C. L., Allington-Smith J. R., 2007, MNRAS, 379, 143
\bibitem[Schmoll, Roth \& Laux(2003)]{Sch2003} Schmoll J., Roth M. M., Laux U., 2003, PASP, 115, 854
\bibitem[Tody(1986)]{Tod1986} Tody D., Proc. SPIE 627, 733 (1986)
\end{thebibliography}
\end{document}